# Franson-Interferometric Bounds on Entangled Two-Photon Absorption

Albin Hedse[1], Sankaran Ramesh[1,2], Luis Matheis[3], Sebastian Gstir[3,4], Andreas Wacker[45], Gregor Weihs[3], Robert Keil[3], Qi Shi,[1] Amitav Sahu,[1] Tõnu Pullerits[1]

[1]Division of Chemical Physics and NanoLund, Lund University, Box 124, 221 00 Lund, Sweden

[2]Centre for Interdisciplinary and Convergent Technologies, Indian Institute of Technology Kharagpur, Kharagpur 721302, West Bengal, India

[3]Department of Experimental Physics, University of Innsbruck, Technikerstraße 25, 6020 Innsbruck, Austria

[4]Epiphany BV, Enschede, Netherlands

[5]Division of Mathematical Physics and Nanolund, Lund University, Box 118, 221 00 Lund, Sweden

## Abstract

Entangled photons offer quantum correlations with no classical analogue. In entangled pair two-photon absorption (ETPA), absorption rate is predicted to scale linearly rather than quadratically with photon flux, promising molecular excitation at fluxes far below the classical threshold. Reported ETPA cross sections nevertheless vary widely across experiments, largely since the observable, the differential of the transmitted flux or weak fluorescence, is difficult to separate from scattering and linear losses. We present a method which uses Franson interference to study entangled two-photon absorption through delay-dependent coincidence measurements on dye molecules. Applying the method to Rhodamine 6G, we observe a small asymmetry in the Franson interference envelope and obtain a model-derived effective cross section of $(2.1 \pm 0.3) \times 10^{-21}\ \mathrm{cm}^2$ for Rhodamine 6G. However, the estimated systematic uncertainty does not allow a definitive ETPA assignment. Instead, the experiment establishes a quantitative bound on the ETPA response and provides a background-free benchmark for future measurements.

## Introduction

Spectroscopy with entangled photons has emerged as an increasingly active research area. By exploiting correlations that have no classical analogue, entangled photons provide access to temporal and spectral information in ways unattainable with conventional light sources, motivating a broad effort to understand how such quantum resources may enhance spectroscopic techniques [1] [2] [3]. Early demonstrations involved interferometry of indistinguishable entangled photons on a single beamsplitter, also known as Hong-Ou-Mandel (HOM) interference [4] [5] [6]. While such approaches highlight the usefulness of quantum correlations, the more substantive advantages associated with entanglement are expected to arise in nonlinear spectroscopic regimes [7] [8] [9]. Although several theoretical proposals outline how nonlinear spectroscopic signals might benefit from entangled light [10] [11], corresponding experimental efforts remain comparatively limited.

Two-photon absorption (TPA) is inherently a nonlinear optical process: because two photons must be absorbed simultaneously, the signal scales with the square of the intensity for light sources producing independent photons [12]. This makes conventional TPA highly inefficient at low intensities and typically requires high laser peak powers that can damage sensitive samples. In entangled two-photon absorption (ETPA), however, the photon pair behaves as a single quantum entity. Consequently, the absorption probability becomes linear in intensity. At low photon fluxes this can dramatically enhance the effective TPA rate, enabling genuinely quantum-enhanced spectroscopic measurements [13] [14].

Realising this promise in practice, however, has proven difficult. The expected ETPA signal at accessible photon fluxes is exceedingly small, so that the observable, typically a decrement in the transmitted pair flux or a weak fluorescence yield, must be extracted from a background many orders of magnitude larger than the effect itself. In transmission geometries the quantity of interest is a fractional reduction in coincidence counts, which is indistinguishable, on inspection alone, from a reduction caused by scattering, imperfect fiber coupling, or any other spectrally uniform linear loss introduced by the sample or its cuvette [15]. Fluorescence detection avoids the differential-transmission problem but introduces its own: hot-band absorption, residual one-photon excitation, and detector afterpulsing can all mimic a genuine two-photon signal at these count rates [16]. Compounding this, the parameters exclusive to ETPA, the entanglement time and entanglement area of the source are critical parameters that set the expected magnitude of the effect but are seldom reported, so that a given cross section cannot readily be transferred between experiments [17]. What is needed, therefore, is not only greater sensitivity but an observable whose functional form distinguishes pair absorption from linear background, independently of the absolute count rate.

ETPA experiments have already been conducted, with one article reporting values for the ETPA cross-section of Rhodamine 6G ranging from $10^{-22}$ $cm^2$ to $10^{-19}$ $cm^2$ [18]. Theoretical upper bounds for ETPA cross-sections have been proposed [17] [19], which often range from $10^{-30}$ $cm^2$ to $10^{-25}$ $cm^2$, in stark contrast with some reported values. Overall, these works have used fluorescence as the key observable to evaluate ETPA. Several concerns regarding previous ETPA measurements have been explored in [17], in addition to the work of determining what a theoretical upper bound to an ETPA cross-section could be.

In this work, we investigate ETPA from another perspective using Franson interferometry [20]. Leveraging the advantages of transmission-coincidence counting, this method enables highly sensitive, background-free measurements of ETPA, providing direct access to the time-dependent absorption response and allowing us to establish quantitative upper bounds on the ETPA cross section in dye molecules. In addition, due to the time-frequency entanglement of the photon pairs used in Franson interference, the time resolution and observable contrast of our method are insensitive to dispersion in the sample and most parts of the setup [21]. We also perform a two-photon fluorescence measurement with classical pulses and estimate the classical TPA cross-section for comparison, providing a bound on the quantum advantage of the ETPA over classical TPA. In our work, we hope to leverage the advantages of transmission-coincidence counting, the most crucial of which is the background-free signal. Unlike in previous measurements of this kind, where the ETPA signal is measured directly as a function of relative photon delay [15], scattering and other single-photon losses do not influence the two-photon measurements presented here.

## Experimental setup

Our setup uses a continuous-wave (CW) single-mode diode laser at 405 nm for the entangled photon pair (EPP) generation, as shown in Figure 1. A λ/2 plate along with a polarizing beamsplitter (PBS) is used to purify the polarization state of the pump beam, in addition to giving an independent control of the pump power. The polarization of the beam is then fine-tuned with one last λ/2-plate just before the lens L1 that focuses the pump beam onto a beta-Barium Borate (BBO)-crystal which is set up for Type-I spontaneous parametric downconversion (SPDC). After the BBO, the remaining pump beam is caught by a beam dump while the down-converted photon pairs are collimated with lenses $L2_A$ and $L2_B$, and each photon channel is filtered by a 650 nm long-pass filter to ensure all pump photons are removed. The photons then pass through band-pass ‘notch’ filters, with a bandwidth of 40 nm centered on 800 nm, before entering fiber collimators which couple the energy-time-entangled photons into optical fibers that take the photons to the subsequent parts of the setup.

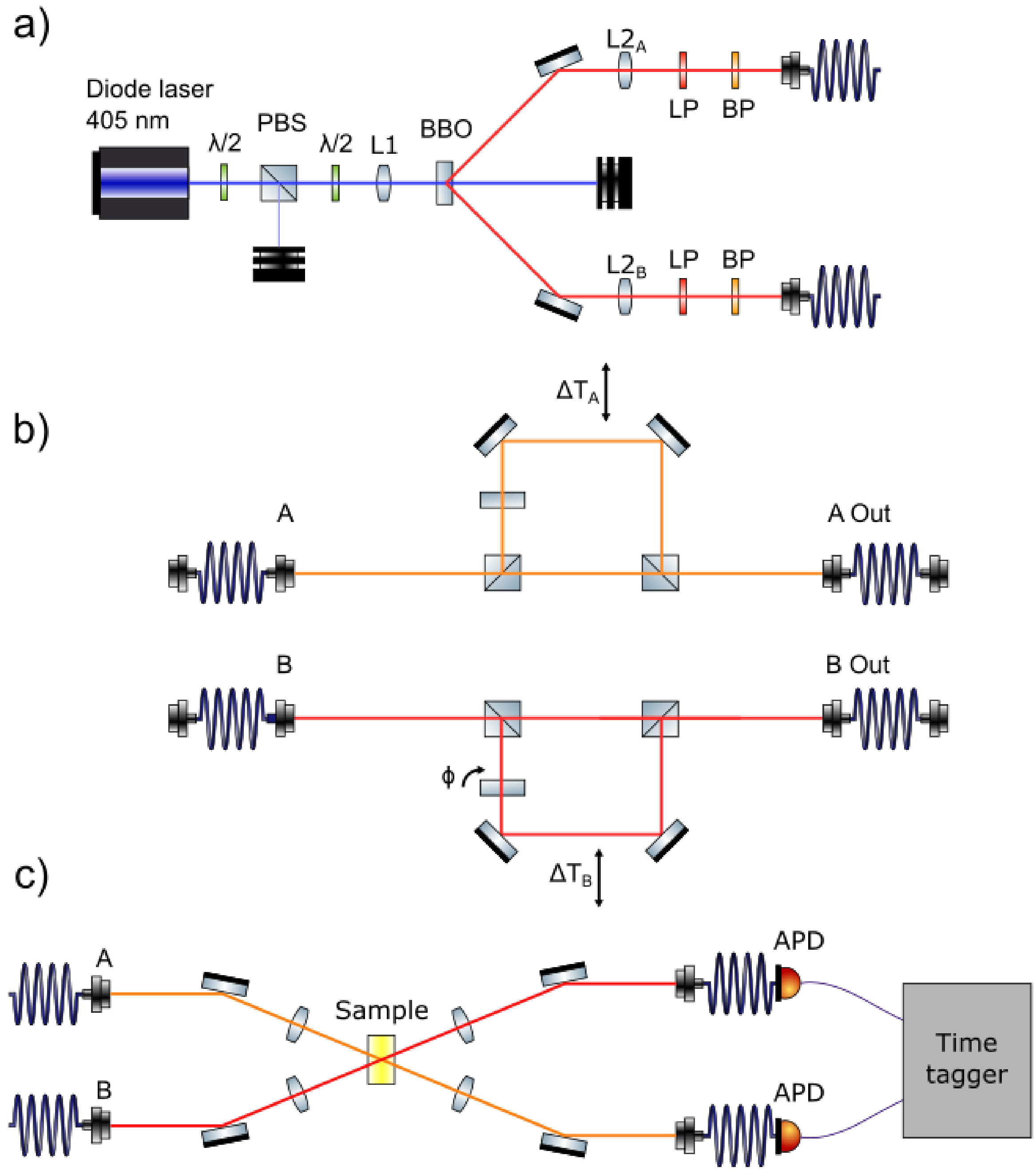

Figure 1: a) The entangled photon pair source. The pump beam, from a single-mode 405 nm diode laser (LM-405-PLR-40-4K Surelock, Coherent), is polarization-filtered with a polarizing beamsplitter (PBS) and the polarization is adjusted with a λ/2 plate. It is then focused by lens L1 on the BBO-crystal (Newlight, NCBBO5200-404(I)-HA6-M12.7, 2 mm thickness) by a lens with a 175 mm focal length. The remainder of the pump beam is caught in a beam dump. The two entangled photons are collimated with lenses $L2_A$ and $L2_B$ and pass a 650 nm long-pass filter (LP) and a 40 nm notch band-pass filter centered on 800 nm (BP) before being coupled out through a fiber collimator. b) The Franson interferometer, with entangled-photon inputs A and B directly linked to the outputs of the EPP source via optical fiber. Out of the two delays, only $\Delta T_B$ is altered throughout the experiment, the 200 μm fused silica phase plate at ϕ is additionally rotated up to 47° as a means of fine-stepping the delay to achieve femtosecond-order delay control. To compensate for the dispersion of the phase plate placed in B, an identical, non-rotating phase plate is placed in the long arm of A as well [22] (The version presented here is simplified, a full version can be found in S1 in the supplementary information, SI). c) The Two-photon absorption portion of the setup. The two output channels of the Franson interferometer coincide on a sample to enable nonlinear interactions between light and sample. Transmitted photons are then collected by fiber collimators and are detected by avalanche photodiodes (Excelitas Technologies, SPCM-NIR-12-FC). A time tagger (Time Tagger Ultra, Swabian instruments) is used for detection of coincidences.

The separate photon channels, from here on referred to as A and B, are then coupled into the Franson interferometer, which consists of two imbalanced Mach-Zehnder interferometers (MZI), one for channel A and B respectively, see Figure 1 b) [23]. Each MZI has two 50-50 beamsplitters, one recombines the photons into joint outputs, of which one is used for each interferometer ('A out' and 'B out'). The long path of each MZI contains a retroreflector on a mechanical step motor and a phase plate consisting of 200 μm of fused silica. The retroreflector is used for coarse tuning the path difference between the two long paths of both interferometers. Fine-tuning is done by rotating the phase plate in channel B while the one in channel A is kept at a fixed angle. This is crucial for observing Franson interference and allows relative delay adjustments in approximately 0.2 fs increments.

To measure ETPA, the outputs of A and B are coupled into a two-photon interaction setup (as shown in Figure 1 c)). Here the photon channels are overlapped at the sample position and measured using the avalanche photodiodes (APD) by coupling them into fibers. The optical paths of both A and B are carefully calibrated, ensuring that photons traveling the short paths of A and B each arrive simultaneously at the sample. This also means that when the path difference between long and short paths in each channel are equal, i.e. $\Delta T_A = \Delta T_B$, long path photons also arrive simultaneously at the sample to one another. This is what offers a method to detect entangled two-photon absorption. The advantages that this brings with respect to other types of ETPA measurements are explained in more detail in the following section.

A camera (Ophir Spiricon SP928 beam profiling CCD camera) in the sample position, in conjunction with an alignment laser (OZ Optics FOSS-21-3S-5/125-780-S-1), is used to ensure maximal spatial overlap of the photons from the different arms at the sample.

## Theory

The advantage of using Franson Interferometry to measure entangled-photon two-photon absorption is that we can study the time-dependence of the absorption on a femtosecond timescale while also tuning the temporal overlap of the photons in the interferometer so that we can exclude any single-photon loss channels. This is due to the main significant single-photon losses remaining constant for all settings of the $\Delta T_B$ -delay, and thus not affecting the delay-dependent change in the measured coincidence counts.

We illustrate our point through a simulation of Franson interference from a simple model based on the derivation in [24], a more detailed version of which can be found in S2 in the SI. The output photon-field of a narrow-band cw-pumped SPDC process is expressed as

$$|\psi\rangle_{\mathrm{SPDC}} = |0_{\mathrm{s}}, 0_{\mathrm{i}}\rangle + \eta \int d\omega_1 d\omega_2 \alpha_{\mathrm{p}}(\omega_1 + \omega_2)\psi(\omega_1, \omega_2)|\omega_{1\mathrm{s}}, \omega_{2\mathrm{i}}\rangle\,, \tag{1}$$

where $\eta$ is a transfer ratio, $\omega_1, \omega_2$ refer to the signal and idler frequencies respectively, $\alpha_{\mathrm{p}}(\omega_{\mathrm{p}})$ refers to the pump field's spectral amplitude where the central frequency of the pump field is $\omega_{\mathrm{p0}}$ and the phase-matching function $\psi(\omega_1, \omega_2)$ defines the spectra of the downconverted photons. Under good experimental conditions, the laser emits a (longitudinally) single-mode beam and $\alpha_{\mathrm{p}}$ can be approximated to be a $\delta$-function, however, this approximation is no longer valid if the laser begins emitting multiple modes. The intensity correlation function for the two photons at their respective detectors in interferometer arms A and B ('A out' and 'B out' in Fig. 1b)), averaged over the pump field is

$$\Gamma_{\mathrm{AB}}^{(2)}(t, t+\tau) = \langle \left| \langle 0 \middle| \hat{\mathrm{E}}_{\mathrm{B}}(t+\tau)\, \hat{\mathrm{E}}_{\mathrm{A}}(t) \middle| \psi \rangle_{\mathrm{SPDC}} \right|^2 \rangle_{\mathrm{p}}\,, \tag{2}$$

where $\hat{\mathrm{E}}_{\mathrm{A}}$, $\hat{\mathrm{E}}_{B}$ are the field operators of the photons at each output of the Franson interferometer, 'A out' and 'B out', respectively, and $\tau$ denotes their detection time difference. Upon integration over this time difference (i.e., the coincidence time window), and assuming a detector time resolution sufficient to distinguish the short and long interferometer arms in both channels, coincidences between mismatched paths are excluded. Under these conditions, the coincidence rate between the two Franson interferometer arms reduces to

$$R_{\mathrm{AB}} \propto 1 + F(\Delta T/2) \left| \gamma_{\mathrm{p}} \left( \frac{\Delta T_{\mathrm{A}} + \Delta T_{\mathrm{B}}}{2} \right) \right| \cos \left[ \frac{\omega_{\mathrm{p0}}(\Delta T_{\mathrm{A}} + \Delta T_{\mathrm{B}})}{2} + \varphi_0 \right]. \tag{3}$$

Here the envelope function $F(\Delta T)$ is defined as the normalized auto-correlation of the Fourier transform of the joint spectral amplitude of the two photons with $\Delta T = \Delta T_{\mathrm{A}} - \Delta T_{\mathrm{B}}$ representing the arrival time difference between the long arms of A and B, see Fig. 1 b). $\Delta T_{\mathrm{A}}$ and $\Delta T_{\mathrm{B}}$ denote the total delay of the long arms of A and B relative to their short arms, respectively, and $\varphi_0$ is a constant phase factor which arises dependent on the beamsplitter output used for the Franson measurement. In the experiment, $\Delta T$ is scanned by varying one of the interferometer delays (here $\Delta T_{\mathrm{B}}$). $\gamma_{\mathrm{p}}$ is the second-order correlation function of the pump field derived from $\alpha_{\mathrm{p}}$, given by:

$$\gamma_{\mathrm{p}} \left( \frac{\Delta T_{\mathrm{A}} + \Delta T_{\mathrm{B}}}{2} \right) = \langle V_{\mathrm{p}}(0)\;\; V_{\mathrm{p}}^{*} \left( \frac{\Delta T_{\mathrm{A}} + \Delta T_{\mathrm{B}}}{2} \right) \rangle / I_{\mathrm{p}}\,, \tag{4}$$

Where $I_{\mathrm{p}}$ is the pump-field intensity, given by the electric field of the pump, $V_{\mathrm{p}}$ which is described as a random classical field:

$$V_{\mathrm{p}}(t) = \frac{1}{\sqrt{2\pi}} \int d\omega_{\mathrm{p}} \alpha_{\mathrm{p}}(\omega_{\mathrm{p}})\, e^{-i\omega_{\mathrm{p}} t}, \tag{5}$$

The form of $F(\Delta T)$ is given by:

$$F(\Delta T) = \exp\left(-\frac{(\sigma * \Delta T)^2}{8}\right), \tag{6}$$

where $\sigma$ represents the (1/e) bandwidth of the downconverted photons in units of frequency. The visibility of the interference fringes is governed by the product $F(\Delta T) * \gamma_{\mathrm{p}}$. Under optimal conditions, the pump laser bandwidth is so narrow that one can ignore $\gamma_{\mathrm{p}}$, however, if this condition is no longer met due to e.g. laser degradation, it quickly has a strong effect on visibility. Visibility here refers to typical interferometric visibility as given by the expression:

$$V = \frac{I_{max} - I_{min}}{I_{max} + I_{min}}, \tag{7}$$

where $I_{max}$ and $I_{min}$ refer to the maximal and minimal intensities in the interference pattern. In the context of Franson interference, the visibility of the fringes is measured the same way but arises directly from the strong correlation between the frequency of the two photons of the down-converted pair. As such, we make direct use of both the time- and frequency correlations of our down-converted photons in our measurement. If we assume the visibility is unity, which can be accomplished in the case that the photons are indeed perfectly frequency-correlated, and ignore the constant phase factor $\varphi_0$ and the correlation function $\gamma_{\mathrm{p}}$, we can simplify the expression in equation (3) to:

$$R_{\mathrm{AB}} \propto 1 + \exp\left(-\frac{(\sigma * \Delta T)^2}{8}\right) \cos\left[\frac{\omega_{\mathrm{p}0}(\Delta T_{\mathrm{A}} + \Delta T_{\mathrm{B}})}{2}\right]. \tag{8}$$

A striking property of Franson interference is its robustness to dispersion: Dispersive effects outside the imbalanced MZIs do not affect the coincidence interference fringes at all, only differential dispersion between the long and short arms of the interferometers can degrade the visibility [21]. This implies that an absorbing sample positioned, as in Fig. 1 c) does not affect the width or visibility of the Franson fringes with its dispersion. In a simple model of absorption, any single-photon absorptivity would result in an even loss of photons across all channels. A sample exhibiting TPA would then cause a reduction in coincidence counts, which depends on the temporal overlap in the long-long case and is constant in the short-short case. Then, TPA of long-path pairs should give a delay-dependent signature in our Franson measurement, as the temporal overlap of these photons only occurs when their path lengths are equal during the scan, giving a built-in TPA "lifetime" measurement characteristic to the scan. Strictly speaking, this is a simplification, as the indistinguishability of long- and short path pairs within the pump coherence length gives rise to the Franson interference, but we keep to the terminology used above here. To model the influence of TPA on the coincidence count scan, we insert a function modeling the temporal characteristic of TPA per sample thickness $d$, assuming it corresponds roughly to the photon overlap and so can be approximated as Gaussian, together with the number density per volume $n$:

$$nd\sigma_{\mathrm{E}}(\Delta T) = nd\sigma_{\mathrm{E0}} \times \frac{\left(1+\exp\left(-\frac{(\sigma*\Delta T)^2}{8}\right)\right)}{2}, \tag{9}$$

where $\sigma_E(\Delta T)$ represents the ETPA cross-section as a function of the delay time in our setup and $\sigma_{E0}$ represents the ETPA cross-section absent any delays between photons, so that $\sigma_{\mathrm{E}}(0) = \sigma_{\mathrm{E0}}$. Equation (9) is then inserted into equation (8), giving:

$$R_{\mathrm{AB}} \propto \left(1 + \exp\left(-\frac{(\sigma * \Delta T)^2}{8}\right)\cos\left[\frac{\omega_{p0}(\Delta T_{\mathrm{A}} + \Delta T_{\mathrm{B}})}{2}\right]\right) * \exp[-\,nd\sigma_{\mathrm{E}}(\Delta T)\,], \tag{10}$$

where the constant component of $nd\sigma_{\mathrm{E}}(\Delta T)$ (that is, $nd\sigma_{\mathrm{E0}}/2$) corresponds to coincidence losses from short-path photon pairs, which should manifest as a constant two-photon loss channel.

A key feature of this approach is its ability to distinguish uniform linear losses from delay-dependent nonlinear absorption. Single-photon losses, such as scattering or imperfect coupling, act independently in each interferometer arm and therefore scale the coincidence rate multiplicatively. Because these processes are insensitive to the scanned delay, they uniformly attenuate the entire Franson interference pattern without modifying its shape. The only exceptions are losses arising from changes in alignment as the delays are changed. These losses are, however, present even without a sample, and so they can be minimized by optimizing the setup's alignment with no sample present (for some data on losses near the sample, see S4 in SI). This has been done carefully in this experiment, such that losses due to changing alignment are negligible here.

To clarify what we mean by this ability to distinguish, we consider three distinct cases. First, in the absence of a sample and assuming negligible losses, the measurement yields an ideal Franson interference pattern with unit visibility in the coincidence counts, while the single-photon counts in each arm remain constant throughout the delay scan (red curve in Fig. 2).

Second, we consider uniform single-photon losses in the setup. Since the delay scan does not affect coupling efficiency or scattering probability over the small path-length variations involved, such losses remain constant during the scan. If the transmissions of arms A and B are reduced to $\eta_A$ and $\eta_B$, the coincidence rate scales as $\eta_A\eta_B$. For example, a 50% loss in each arm reduces the coincidence rate by approximately 75%. Importantly, this attenuation applies uniformly to the entire interference pattern: the maxima, minima, and baseline are reduced by the same multiplicative factor. Consequently, single-photon losses lower the overall count rate but do not introduce any delay-dependent features. Provided the coincidence rate remains sufficiently high for reliable statistics, the interference structure itself is unaffected.

Third, we consider two-photon losses. These fall into two categories. Delay-independent two-photon absorption arises from photon pairs that take the short paths in both interferometers and therefore overlap at the sample for all delays. This produces a constant reduction in coincidence counts, and is described in the constant term of eq. (9).

In contrast, delay-dependent two-photon absorption occurs only when photons traversing the long arms of both interferometers overlap temporally at the sample. This overlap is restored only near zero relative delay and therefore produces a localized suppression of the interference maxima. Such losses modify the shape of the interference envelope rather than merely its overall amplitude. At the photon fluxes used here, other sample-induced nonlinear optical processes are expected to be negligible, making two-photon absorption the relevant nonlinear mechanism capable of producing such a delay-dependent coincidence loss. Instrumental effects, including residual dispersion and delay-scan reconstruction artifacts, can nevertheless also modify the interference envelope and are considered separately below. The simulated signatures of different ETPA cross-sections are illustrated by the blue and green curves in Fig. 2. For a more detailed illustration of the different cases (pure interference, single-photon absorption, two-photon absorption with/without single-photon absorption), see section S7 in the SI.

The distinction between uniform attenuation and structural modification is central to our method. Single-photon losses reduce the total coincidence rate but leave the delay dependence unchanged, whereas delay-dependent two-photon absorption alters the structure of the Franson interference pattern. By monitoring changes in the envelope near zero delay, we isolate nonlinear two-photon effects from background linear losses. In the absence of a measurable delay-dependent modification, the experiment therefore establishes an upper bound on the ETPA cross-section.

The trade-off of the Franson geometry is a reduced absolute coincidence flux as a full ¾ of coincidences are lost due to the outcoupling of the photon pairs. However, this reduction is compensated by background-free coincidence detection in intrinsically separated signal and idler channels, together with independent control of the relative photon delay. To our knowledge, this combination has not been demonstrated previously in transmission-mode ETPA experiments, which typically rely on statistical separation of photons at a beamsplitter rather than intrinsic channel separation from source to detection [18] [16].

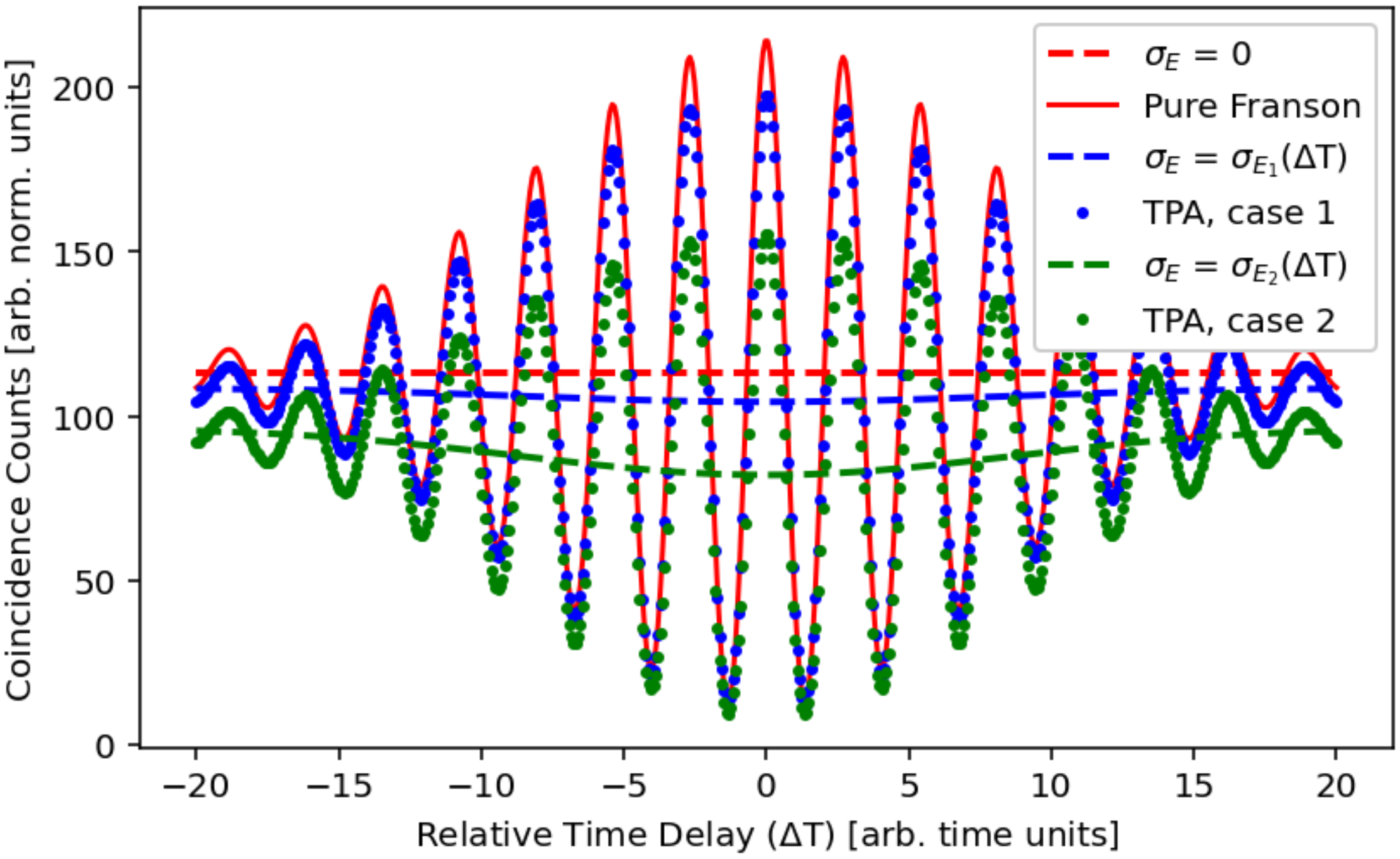

Figure 2: A simulation of Franson interference fringes in case of visibility being unity with no sample present (red, solid), the same fringes except for the presence of a sample with an ETPA cross-section of approximately $3*10^{-21}$ $cm^2$ in the sample position (blue dots, "case 1") and again the same Franson interference pattern with a sample, with an ETPA cross-section $1*10^{-20}$ $cm^2$, in the sample position (green dots, "case 2"). The dashed line patterns represent coincidence counts without Franson interference. The three cases displayed are: without any sample, with the sample from case 1 and with the sample from case 2, shown as red, blue and green, respectively.

Our proposed advantages of using Franson interference for measuring TPA rely on a few key points: That our setup is properly aligned so that the long-path temporal overlap indeed occurs at $\Delta T = 0$ and that we by that same token have constant overlap between short-path photons throughout the measurement. This was achieved by carefully measuring the relative photon arrival times by adjusting the delay between interferometer arms A and B and performing multiple coincidence measurements to maximize the temporal photon pair overlap at the sample position. In addition, we assume that any TPA will affect the coincidence counts linearly, i.e. that the interference itself is unaffected by any two-photon losses. This is justified since Franson interference is a nonlocal effect and is in general not affected by anything after the outputs of the two MZI's, provided enough output photons can still be measured.

## Results

Measured Franson interference traces with two samples, Rhodamine 6G and Coumarin 152, are shown in Fig. 3 (steady state spectra shown in section S3 of the SI). The data collection time was 18 hours, during which we require stability of the output mode of the pump laser. Experiments were performed with a pump power of 13.5 mW. The laser power was optimized to allow stability over the whole measurement while maintaining a sufficiently high photon count for good quality interference fringes, see SI.

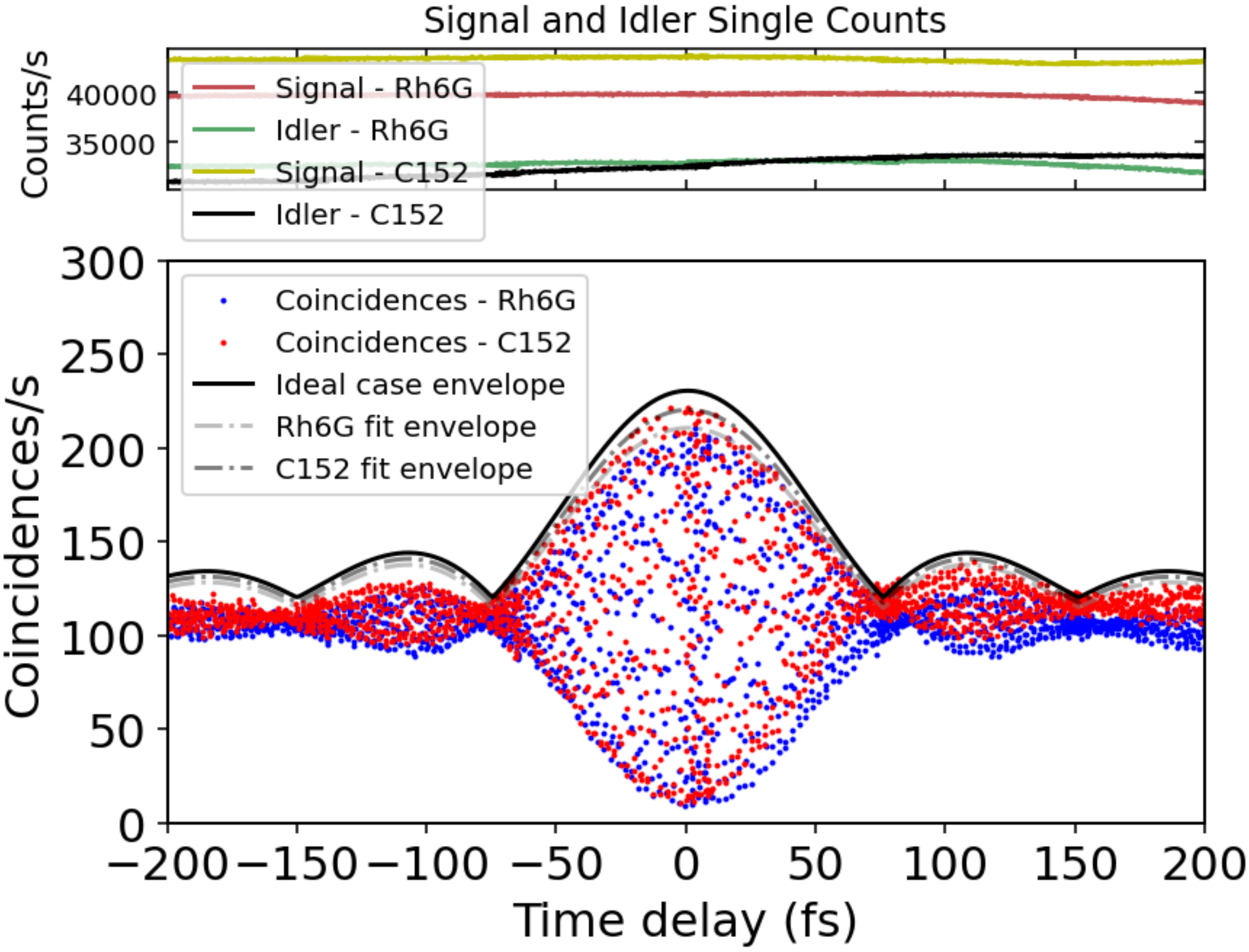

Figure 3: Top plot: The counts in the above plot represent the total photon counts in each channel. Bottom plot: The lower plot represents the coincidences counted between the two detection channels. Franson interference patterns measured with a Rhodamine 6G sample, in blue, or Coumarin 152, in red, in the sample position, as described in Fig. 1 c). The envelopes displayed are derived from the absorption model described in eq. (10). The cases being modeled are as follows: Simulated Franson interference envelope without a sample (black line) with visibility assumed to be in line with that of the sample (modeled as 92 % visibility), the absorption model as applied to Rh6G (light grey, dash-dotted) and as applied to C152 (dark grey, dash-dotted). The Franson interference patterns for Rhodamine 6G and Coumarin 152 have visibilities of 90.6 and 91.6 % respectively, indicating a very strong spectral correlation between the detected photon pairs. Measurements are both taken at a laser output power of 13.5 mW. Precise coincidence counts, both measured and modeled, are shown in Table 1 below. A closer view of the experimental data and modeled fit can be seen in Figs. 4 and 5.

For a closer look at the interference counts and the fit of the absorption model to the experimental data, see Figures 4 and 5 below. Table 1 summarizes the most relevant data for evaluating the model: the minimum and maximum coincidence counts at the central Franson fringe, as well as the mean coincidence counts outside the fringes, together with the corresponding model values.

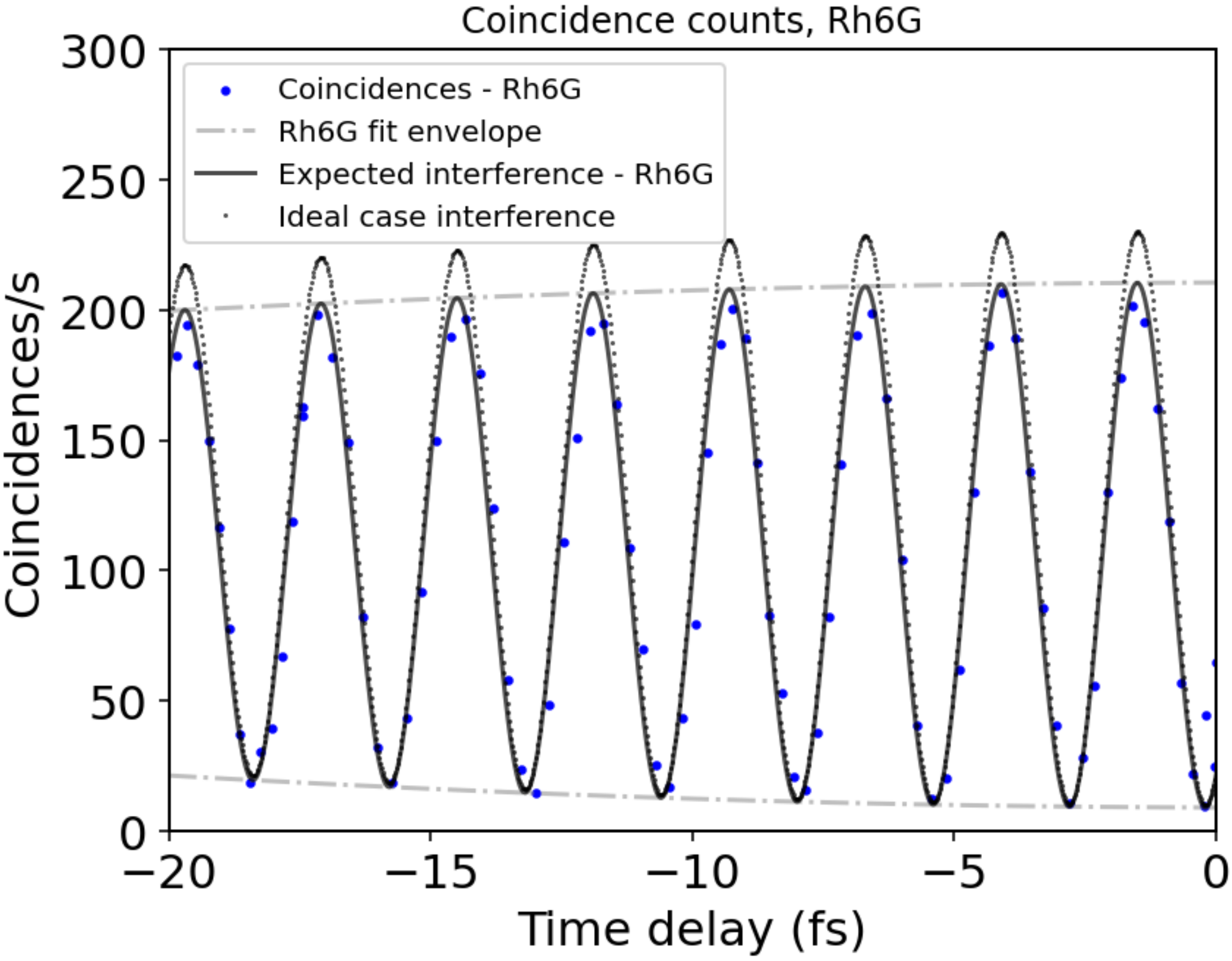

Figure 4: The coincidence count data for Rhodamine 6G along with the modeled Franson fringes with the ETPA absorption model applied in a window near zero delay where the time-dependent component of the ETPA is expected to have the largest effect on the coincidence signal. Experimental Rh6G coincidence counts are shown as blue dots, with the envelope given by the model in the grey dash-dotted line and the modeled interference with phase fitted to the data given by the black line. The black dots represent the ideal case envelope from Fig. 3 applied to the interference, i.e. what the interference pattern would look like without any ETPA in the model. The interference minima in the two cases are similar because they are, in absolute numbers of coincidences/s, less affected by ETPA than the maxima.

Using the model in Eq. 13 and the data in Table 1, we obtain fit-derived effective ETPA cross-sections of $\sigma_{E0} = (2.1 \pm 0.3) \times 10^{-21}\ \mathrm{cm}^2$ for Rh6G and $\sigma_{E0} = (1.1 \pm 0.3) \times 10^{-21}\ \mathrm{cm}^2$for C152, when only the standard deviation of the coincidence counts is included as uncertainty. For Rh6G, the fitted value is of the same order as that reported in Ref. [18], although the sample concentrations differ by approximately a factor of three. The model reproduces the maxima and minima of the interference patterns reasonably well, as seen in Figs. 4 and 5 and Table 1, but it does not account for the baseline mean coincidence counts within the measurement uncertainty alone, which is about 3 counts/s for both samples (see Table 1). This is further complicated by the absence of a control measurement of pure solvent acquired under directly comparable experimental conditions (see S5 in the SI for more details), which would have helped constrain the minimum and maximum counts more reliably. While ETPA should not, in principle, alter the visibility through a purely linear loss channel, a control measurement would nevertheless be valuable for verifying this assumption experimentally.

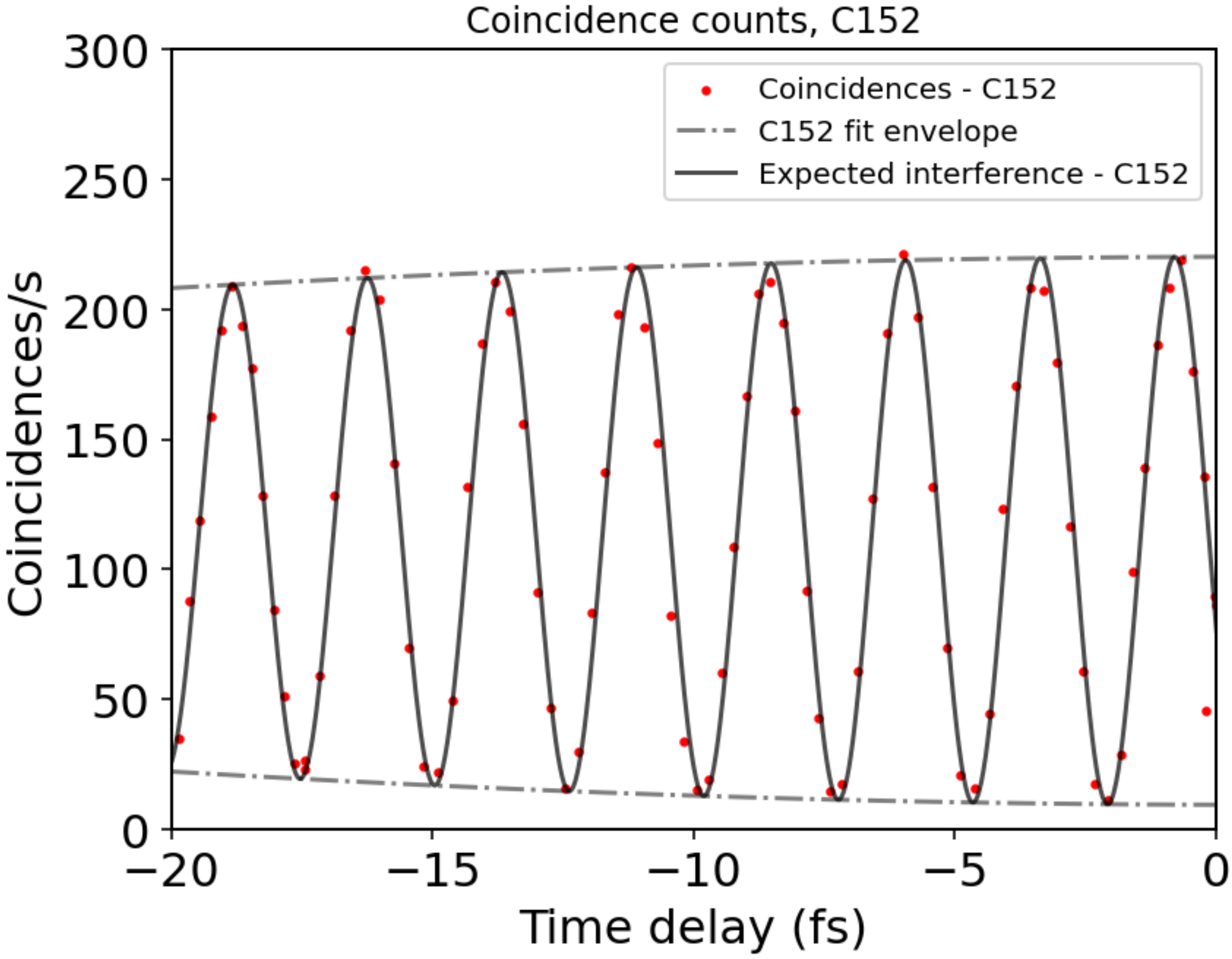

Figure 5: The coincidence count data for Coumarine 152 along with the modeled Franson fringes with the ETPA absorption model applied in a window near zero delay where the time-dependent component of the ETPA is expected to have the largest effect on the coincidence signal. Experimental C152 coincidence counts are shown as red dots, with the envelope given by the model in the grey dash-dotted line and the modeled interference with phase fitted to the data given by the black line.

As shown in Fig. 3, the Franson interference pattern exhibits a slight asymmetry along the delay axis (manifesting in the different visibilities of the side peaks), indicating residual odd-order group velocity dispersion mismatch between the two interferometer arms, or possibly third-order contributions from phase plates and mirrors [21] [25]. This reduces the fringe visibility below 100% and, without an independent control measurement of pure solvent, its exact magnitude cannot be determined. In addition, the finite time resolution of the detectors also limits the visibility. Together, these effects introduce uncertainty in the fringe contrast and therefore in how confidently the observed asymmetry can be attributed to ETPA. Further uncertainties arise from details of the scan reconstruction and from measurement-to-measurement normalization. Taken together, these contributions correspond to an estimated uncertainty in the extracted cross-section of $(0.6\text{–}1.3) \times 10^{-21}\ \mathrm{cm}^2$ (see S8 in the SI for details).

Several additional technical issues limit the accuracy of these results: Firstly, small changes in the longitudinal laser output mode can make otherwise similar measurements difficult to compare (For examples, see S5 in the SI). Secondly, the beam-direction drift can alter the SPDC conditions and the fiber outcoupling efficiency, leading to asymmetric changes in the signal and idler counts. Thirdly, the band-pass filtering used in the source also increases the effective entanglement time (by reducing the time-correlation between the photon pair). This makes a large quantum advantage less likely unless the entanglement area (the spatial correlation between the photons in the pair) is substantially different to our estimations, but due to our setup this is effectively replaced by the interaction area, this is highly unlikely. These parameters are difficult to compare directly because they are rarely reported, although our estimates are comparable to the few values available in the literature [15,17]. Finally, it would have been desirable to perform measurements at different pump powers as well, as it would enable us to directly compare the scaling of ETPA with increases in entangled photon flux. However, we were unable to measure at multiple pump powers because of laser stability issues (for details, see S6 of the SI).

Together with the discrepancy between the fitted cross-section and the theoretical upper bound discussed below, these limitations make it difficult to attribute the small difference between C152 and Rh6G in Fig. 3 unambiguously to ETPA. The observed odd-order dispersion further suggests that the Franson fringes may not be perfectly symmetric even in the absence of a sample, which would be clarified by a pure solvent control measurement.

Table 1: Values from the data displayed in Figs. 3,4 and 5. Standard deviation of coincidence counts at baseline is calculated from the experimental variance of the baseline, which is roughly Poissonian. Maximum and minimum here refers to the counts at the central peaks and valleys in the coincidence interference pattern for the experimental data and the maximum and minimum given by the absorption model in the 'model fit' cases. Baseline coincidences refer to the mean coincidence counts outside the Franson interference fringes, in the model case the counts at these points are identical to the counts at the nodes in the envelope, which should correspond to the mean counts in the measurement absent any asymmetry due to delay-dependent TPA.

| **Sample** | **Maximum coincidence counts/s** | **Minimum coincidence counts/s** | **Baseline coincidence counts/s** | **Standard dev. of coincidence counts/s (Baseline)** |
|---|---|---|---|---|
| **Rh6G** | 210.4 | 9.3 | 104.5 | 2.7 |
| **Model fit Rh6G** | 210.6 | 8.8 | 114.4 | N/A |
| **C152** | 221.3 | 10.9 | 114.5 | 3.0 |
| **Model fit C152** | 220.3 | 9.2 | 117.3 | N/A |

A definitive result would require a substantially higher single-mode pump power to improve the signal-to-noise ratio, here limited by the statistical variation in the coincidence counts at each delay point. In its current state, however, the laser cannot provide such flux in a sufficiently stable manner. Moreover, the Franson interferometer is still too dispersive to exploit the full bandwidth of the downconverted photons as there are still third-order dispersion contributions from inside the MZI. In addition, the SPDC source is coupled through optical fibers of unequal lengths (estimated to be a few cm), which will contribute as well. As a result, we cannot readily increase the time resolution of the setup by removing the band-pass filters. If TPA is indeed present, reducing the individual-photon coherence time would increase the TPA signal and thereby strengthen the case for its identification. Taken together with the estimated uncertainty of $(0.6–1.3) \times 10^{-21}$ cm$^2$, these limitations make it difficult to claim a definitive observation of ETPA, although the results remain a compelling proof of concept.

Table 2: Comparison of reported ETPA cross-sections in similar studies. Some experiments yielded a null result, for these, the estimated highest upper bound is shown. In our case (top row), the measurement should rather be treated as an upper bound (see the discussion of uncertainty) as opposed to a claim of a confirmed measurement of ETPA.

| **Sample** | **Concentration (mmol/L)** | **Entanglement time $T_E$ (fs)** | **Entanglement area $A_E$ (µm$^2$)** | **ETPA cross-section $\sigma_{E0}$ (cm$^2$)** |
|---|---|---|---|---|
| **Rh6G [This work]** | 0.16 | 100 | 7-500 | $2.1 \pm 0.3 * 10^{-21}$ |
| **Rh6G [17]** | 1.5 | 1620 | 2.1-13700 | N/A ($10^{-25}$ est.) |
| **Rh6G [18]** | 0.038 | 140 (est.) | N/A | $1.9 \pm 0.9 * 10^{-21}$ |
| **Rh6G [18]** | 4.5 | 140 (est.) | N/A | $9.9 \pm 4.9 * 10^{-22}$ |

## Comparison with classical two-photon absorption and ETPA upper bound

Regarding the relative contributions of entangled photon pair absorption and classical two-photon absorption to the total two-photon absorption we use the model for two-photon absorption rate for a pulsed source as described in [17]:

$$R_C = \frac{\sigma_C \mu^2 g^{(2)}}{2T^2A^2} \tag{10}$$

Where $R_C$ is the classical TPA rate, $\sigma_C$ is the classical TPA cross-section, $\mu$ is the mean photon number, $g^{(2)}$ is the second-order photon correlation function, here $T$ is the pulse duration and $A$ is the beam area. Each SPDC emission event is statistically a thermal state, however, as the ETPA absorption event relies on two separate photons we use the cross-correlation between the signal and idler photons of a pair in the place of the classical second-order correlation. For a degenerate Type-I SPDC process, the cross-correlation can be modeled as a single-mode squeezed vacuum which has the second-order correlation function:

$$g^{(2)} = 3 + \frac{1}{\mu} \tag{11}$$

Which gives, when substituted in place of the second-order correlation function for the SPDC photon pairs:

$$R_E = \frac{1}{2}(\sigma_{\mathrm{E}}\phi + 3\sigma_{\mathrm{C}}\phi^2). \tag{12}$$

Where $R_{\mathrm{E}}$ is the rate of TPA for a given sample, $\sigma_E$ is the ETPA cross-section, $\phi = \frac{\mu}{TA}$ is the (entangled) photon flux. $\sigma_{\mathrm{E}}$ is defined as:

$$\sigma_{\mathrm{E}} \approx \frac{\sigma_C}{T_{\mathrm{E}}A_{\mathrm{E}}}. \tag{13}$$

Provided the photons generated in the experimental setup are more closely correlated in time than the coherence time of the pump photons, the above expression holds. One can then use the entanglement time $T_{\mathrm{E}}$ which is a measure of the time-correlation between the signal and idler photons of a pair, which depends on the entangled-photon source, which includes the laser, the SPDC geometry, etc. The entanglement area $A_{\mathrm{E}}$ is a measure of the spatial correlation between the two photons which also depends on focusing geometry of the pump. In practice, however, after passing through the experimental setup, this correlation is effectively destroyed and replaced with the interaction region in the sample, which is here determined by the spatial overlap of the two beams. Provided these parameters remain static, the photon flux at which entangled-photon absorption contributes equally to the classical two-photon absorption rate is fixed regardless of the two-photon cross-section of the material. In reality, it is not likely that the cross-correlation follows the single-mode squeezed vacuum so closely, which will affect the coefficients in the above equations. This can change the balance of the entangled/classical components at a given flux, but should not change the linear/quadratic scaling [17]. For our setup, we have visualized a 'crossing point', i.e. the photon flux where classical TPA and ETPA contributions are equal, for our setup with the single-mode assumption in Figure 6, which we calculate by setting $\sigma_E\phi = 3\sigma_C\phi^2$ and solving for $\phi$. For reference, our coincidence counts of approximately 100 per second would appear at 2 * $10^{11}$ $cm^{-2}$ $s^{-1}$, many orders of magnitude below this crossing point, which means that any potential contribution from classical TPA is completely negligible. Accounting for losses in our setup and detector inefficiencies, the actual coincidence counts at the sample position are around 650 per second (see S8 in the SI for details), a flux of approximately 1.2 * $10^{12}$ $cm^{-2}$ $s^{-1}$. Even accounting for those detector inefficiencies and losses that cause coincidence losses proportional to the square of our single-photon losses, we are still far off from reaching this crossing point. With the spot size in our case being 25 μm in diameter, this would require 2.3 * $10^{14}$ photons per second in to reach the equivalence point in our setup, corresponding to 57 μW of entangled photon power, which is well beyond our capabilities, which currently lies in the femtowatt regime. It has been noted by Parzuchowski et al. [17] that the reported quantum enhancement requires entanglement areas well below what is seemingly possible to achieve, with values of the order of $10^{-9}\mu m^2$. This leads us to the necessity for standardization with regards to entangled two-photon absorption measurements.

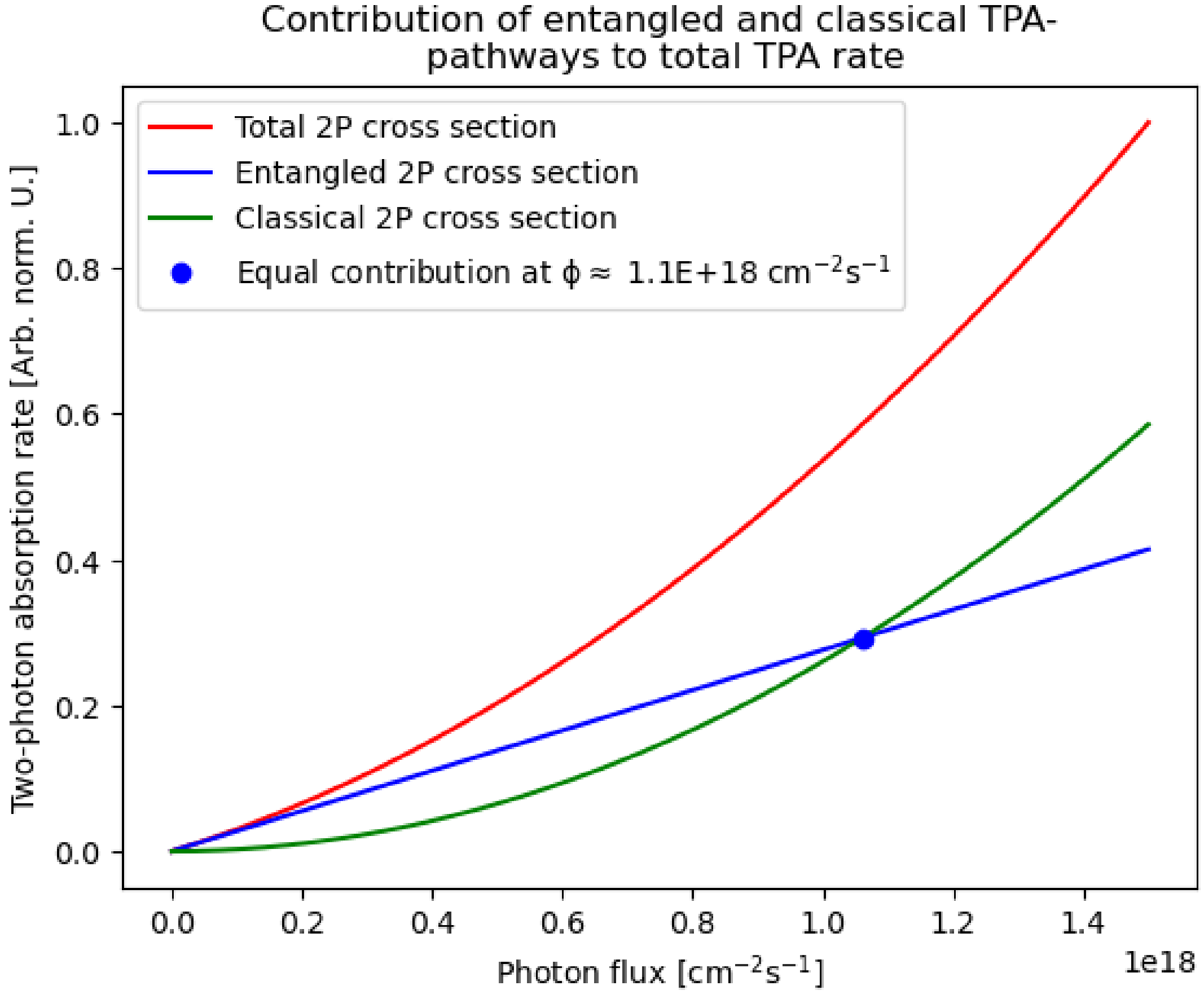

Figure 6: A simulation of the relative magnitudes of two-photon absorptions near the photon flux where classical two-photon (green) absorption 'overtakes' the entangled-photon absorption's contribution (blue) to the total two-photon absorption rate (red). Our experimental photon flux is around 6-7 orders of magnitude less than this point.

A key point is that the measurement-critical parameters of "entanglement time" and "entanglement area" as termed by Parzuchowski et al. [17] are not widely documented or accounted for in studies of this kind. This makes it difficult to avoid "apples-to-oranges" comparisons between experiments in different laboratories, barring the few cases where the conditions of entangled-photon generation, focusing optics, and detection setups are similar enough to justify straightforward comparison.

For comparison, we have included measurements of classical two-photon excited fluorescence under classical pulsed excitation of a Rhodamine 6G sample in both ethanol and methanol, in addition to the Coumarin 152-sample, seen in Fig. 7. The classical measurement was carried out with a pulse width of approximately 20 fs, with a wavelength maxima at 790 nm on the same samples as the ETPA measurements shortly after ETPA was measured. Interestingly, from our data we extract a factor 3 difference in the classical TPA cross-section, not the same as the factor of two that our absorption model suggested between the ETPA cross-sections, but not orders of magnitude different either.

We have used the classically measured values to compare with the literature and thereby calculate our setup's estimated quantum advantage. Based on an entanglement time $T_E \approx 100$ fs and area between 7-490 μm$^2$ with our best estimate at $A_E \approx 400$ μm$^2$ from the interaction area itself, we estimate the entangled TPA cross-section in our case to be around 6.5* 10$^{-30}$ cm$^2$, using equation 13 above [17]. We believe our estimates to be accurate and realistic with respect to the capabilities of our setup. Given the uncertainties in the entanglement time and area, the precise numerical discrepancy should not be overinterpreted; nevertheless, its magnitude strongly suggests that the fitted envelope distortion cannot be assigned straightforwardly to molecular ETPA. Rather, it reinforces the conclusion that systematic effects presently determine the experimentally accessible bound.

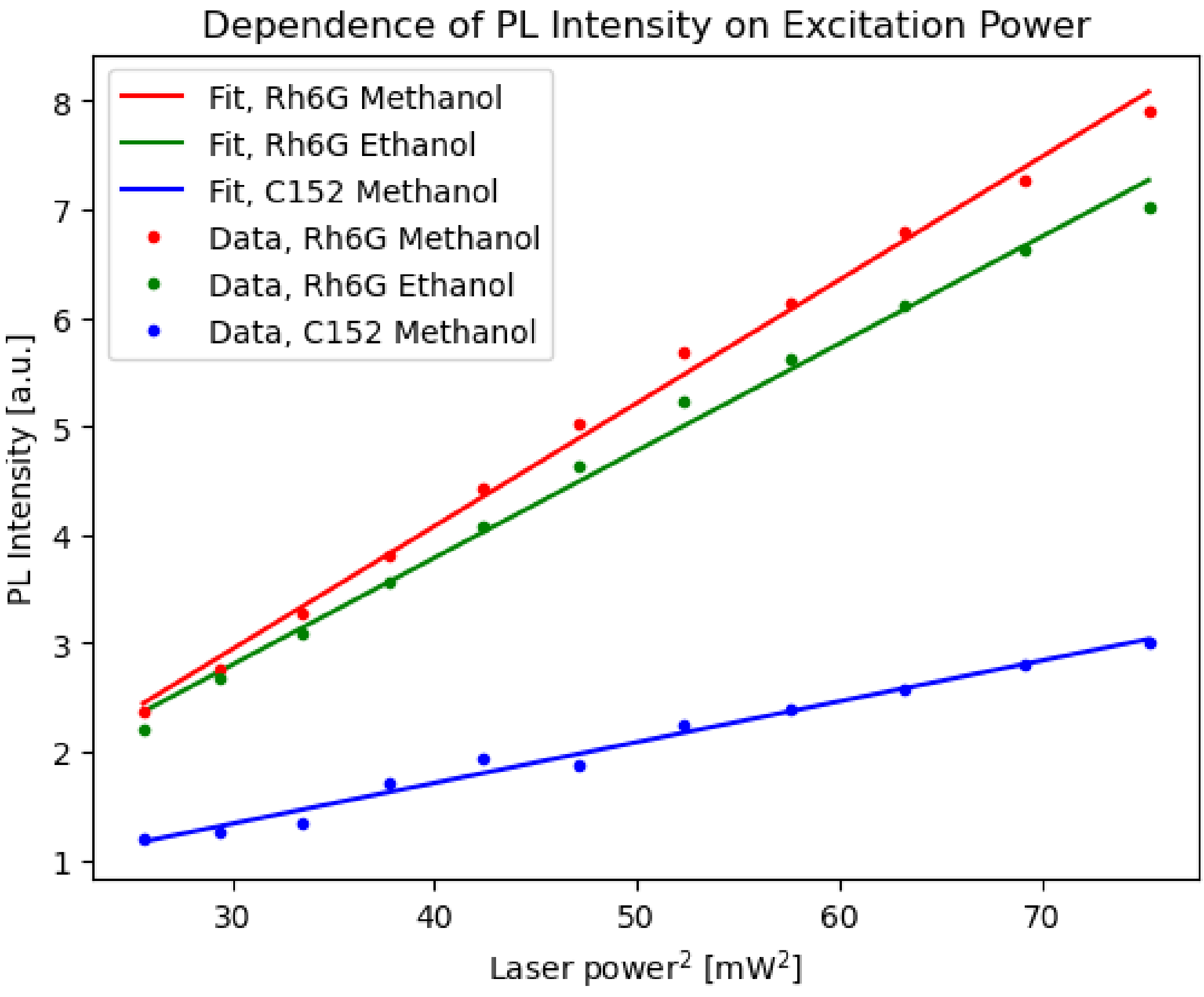

Figure 7: Measurements of photoluminescence from non-entangled two-photon absorption with a pulsed 790 nm laser. The Rh6G samples are of similar concentration, so any difference is mostly attributable to the different solvents. Absorption profiles are not significantly different between the samples, so slope differences are attributable mainly to solvents within margin of error of the relevant cross-section calculation.

Conclusion

We propose an interferometric setup for ETPA measurements where Franson fringes provide additional contrast. The noise floor is very low due to our background-free collection, limited mainly by the statistical uncertainty in pair collection. At our photon fluxes and time resolution, the accidental coincidence rates arising from the simultaneous generation of multiple photon pairs are many orders of magnitude below the statistical pair generation uncertainties and so are not considered to be of a significant impact here. Another advantage of our setup is that we can make use of both the time- and frequency correlations of the generated entangled photons, fully harnessing the strong correlations that the photon entanglement yields.

Our set-up was tested for Rh6G and C152, and the measured Franson interference traces were analyzed using the model developed here. For Rh6G, the fit yields an effective ETPA cross-section of $2.1 \pm 0.3 * 10^{-21}$ $cm^2$ when only the statistical uncertainty of the coincidence counts is considered. This value is comparable to previously reported experimental values [18], but exceeds theoretical estimates based on the source and molecular parameters by approximately nine orders of magnitude [17]. Moreover, the estimated systematic uncertainty of the present experiment is comparable to the fitted value itself. In particular, residual dispersion, scan reconstruction, and source instability cannot be separated reliably from a possible ETPA-induced modification of the Franson envelope in the absence of a directly comparable pure-solvent control measurement. We therefore do not interpret the fitted value as a definitive measurement of ETPA. Instead, the present results establish the sensitivity and corresponding upper-bound scale accessible with our implementation and demonstrate the applicability of Franson interferometry as a method for constraining entangled two-photon absorption. Interferometry for the purpose of measuring ETPA is an area which has only been touched on lightly in literature [26]. We believe that the field in general, and Franson interferometry in particular, can be of benefit to those wishing to explore ETPA and other nonlinear light-matter interactions further.

## Acknowledgments:

We thank Michael Schlosser for fruitful discussions. This research was funded by Royal Physiographic Society of Lund, Olle Engkvist foundation grant 235-0422, Swedish Foundation of Strategic Research grant IS24-0005, KAW WACQT program, the Austrian Science Fund (FWF), projects 10.55776/F71, 10.55776/P30459, 10.55776/I2562.